\documentclass{article}
\usepackage{epsfig,amsmath,amssymb,amsthm,slashed}
\usepackage{graphicx}
\usepackage{color}
\usepackage{siunitx}
\usepackage{color}
\usepackage{bm}
\usepackage{subfig}
\newtheorem*{prop}{Proposition}

\newcommand{\di}{{\rm d}}

\newcommand{\tr}{{\rm tr}}  
  
\newcommand{\ee}{{\rm e}}

\newcommand{\lieb}{{\cal L}_\beta}

\def\wT{{\widehat T}}
\def\wj{{\widehat j}}
\def\wJ{{\widehat J}}

\def\wP{{\widehat P}}

\def\wQ{{\widehat Q}}

\def\wrho{{\widehat{\rho}}}

\newcommand{\beq}{\begin{equation}}
\newcommand{\eeq}{\end{equation}}                                                                               
\newcommand{\bea}{\begin{eqnarray}}
\newcommand{\eea}{\end{eqnarray}}

\begin{document}
\begin{center}
\begin{Large}
{\bf Thermodynamic equilibrium in relativity: \\
four-temperature, Killing vectors and Lie derivatives}
\\
\vspace*{1cm}
\end{Large}
F. Becattini \footnote{becattini@fi.infn.it}\\
\medskip
{\it Department of Physics and Astronomy, University of Florence, Italy}
\end{center}
\vspace*{1cm}
The main concepts of general relativistic thermodynamics and general relativistic 
statistical mechanics are reviewed. The main building block of the proper relativistic 
extension of the classical thermodynamics laws is the four-temperature vector $\beta$, 
which plays a major role in the quantum framework and defines a very convenient 
hydrodynamic frame. The general relativistic thermodynamic equilibrium condition 
demands $\beta$ to be a Killing vector field. We show that a remarkable consequence 
is that all Lie derivatives of all physical observables along the four-temperature 
flow must then vanish.
\vspace*{1.5cm}

\section{Introduction}

Relativistic thermodynamics and relativistic statistical mechanics are nowadays 
widespreadly used in advanced research topics: high energy astrophysics, cosmology, 
and relativistic nuclear collisions. The standard cosmological model views the 
primordial Universe as a curved manifold with matter content at (local) thermodynamic
equilibrium. Similarly, the matter produced in high energy nuclear collisions is
assumed to reach and maintain local thermodynamic equilibrium for a large fraction
of its lifetime. 

In view of these modern and fascinating applications, it seems natural and timely to 
review the foundational concepts of thermodynamic equilibrium in a general relativistic 
framework, including - as much as possible - its quantum and relativistic quantum field 
features. I will then address the key physical quantity in describing thermodynamic 
equilibrium in relativity: the inverse temperature or four-temperature vector $\beta$. 
I will show how this can be taken as a primordial vector field defined on the sole 
basis of thermodynamic equilibrium and ideal thermometers, and its outstanding 
geometrical features in curved spacetimes.

\section{Entropy in relativity}
\label{entrel}

The extension of the classical laws of thermodynamics to special relativity raised
the attention of Einstein and Planck themselves \cite{einstein,planck}. Their viewpoint
is still the generally accepted one, with a later alternative approach put forward in 
the '60s \cite{ott,kampen}, which will be discussed in Sect~\ref{thermom}.
 
The first question, when trying to extend classical thermodynamics to relativity, 
is how to deal with entropy. More specifically, should entropy be considered as a 
scalar or the time component of some four-vector, like energy? 
The non-controversial answer is that total entropy should be taken as a relativistic 
scalar, for various well-founded reasons. Here, I will present a general relativistic
argument based on the second law of thermodynamics, that is total entropy of the 
universe must increase in all physical processes. As in some finite portion of the 
spacetime entropy may decrease, at it is borne out by our daily experience, the only 
sensible choice for an extensive non-decreasing quantity is the result of an integration. 
Since in general relativity an integral can only be a scalar to be generally covariant, 
entropy must then be a scalar. Examples of integral scalars are well known: the 
action, which is the integral over a finite four-dimensional region of spacetime of 
the lagrangian density:
$$
   A = \int_\Omega \di^4 x \; \sqrt{-g} \cal{L} \; ;
$$
the total electric charge, which is the integral over a 3D spacelike hypersurface
$\Sigma$ of a conserved current:
$$
   Q = \int_\Sigma \di \Sigma \; n_\mu j^\mu  \; .
$$
where $n$ is the (timelike) normal unit vector to $\Sigma$ and $\di \Sigma$ its 
measure. Similarly, total entropy should result from the integration over a 3D 
spacelike hypersurface of an entropy current $s^\mu$:
\beq\label{entropy}
   S = \int_\Sigma \di \Sigma \; n_\mu s^\mu 
\eeq

Even if this approach is apparently the most reasonable relativistic extension,
it should be pointed out that the total entropy (\ref{entropy}) is meaningful 
only if entropy current is conserved, i.e. if $\nabla_\mu s^\mu = 0$, which applies 
only at global thermodynamic equilibrium. In a non-equilibrium situation 
$$
 \nabla_\mu s^\mu \ge 0
$$
and the total entropy will depend on the particular hypersurface $\Sigma$ chosen.
Otherwise stated, in non-equilibrium the total entropy is an observer-dependent 
quantity as two inertial observers moving at different speed have two different 
simultaneity three-spaces. Only if $\nabla_\mu s^\mu = 0$, because of the Gauss' 
theorem, the total integral $S$ in eq.~(\ref{entropy}) is independent of $\Sigma$ 
provided that the entropy flux at some timelike boundary vanishes. If $\Sigma$ is 
a hypersurface at some constant time, however the time is defined, this also implies 
that total entropy will be time-independent: precisely our familiar classical 
definition of equilibrium.

\section{Temperature and thermometers in relativity}
\label{thermom}

The first physical quantity encountered in thermodynamics textbooks is temperature. 
It is then natural to wonder how relativity affects the classical temperature notion. 
There has been a long-standing debate about the way temperature changes with respect 
to Lorentz transformations (see e.g. \cite{arabi} for recent summary). The debate 
stemmed from the possible ambiguity in the extension the well-known thermodynamic 
relation (at constant volume):
\beq\label{classic}
  T \di S = \di U
\eeq
If this is seen as a scalar relation, one would most likely conclude, like Einstein 
and Planck \cite{einstein,planck}, that $\di U/T$ must be generalized to be a scalar 
product of four-vectors $\beta = (1/T) u$ (see later on) and $\di P$, being $u$ the 
four-velocity of the observer and $P$ the four-momentum:
\beq\label{ep}
  \di S = \frac{1}{T} u \cdot \di P
\eeq
Conversely \cite{ott,kampen}, if the relation (\ref{classic}) is seen as the time 
component of a four-vectorial relation, with $\di U = \di P^0$, then one would 
accordingly conclude that $T$ is the time component of a four-vector $T^\mu = T 
u^\mu$ and:
\beq\label{ovk}
 T u^\mu \di S = \di P^\mu
\eeq
These two different extensions of the classical thermodynamic relation involve two
converse answers to a relevant physical question: what does a moving thermometer - with respect 
to the system which is in thermal contact with - measure? Or, tantamount, what does 
a thermometer at rest measure if it is put in thermal contact with a moving system 
with four-velocity $u$? It should be stressed that here by thermometer we mean an 
idealized gauge with zero mass, pointlike and capable of reaching equilibrium instantaneously 
(zero relaxation time) with the system which it is in contact with. In the first 
option, the temperature measured by a thermometer at rest in a moving system is 
{\em smaller} by a factor $\gamma$, in the latter case is {\em larger} by a factor 
$1/\gamma$, where $\gamma$ is the Lorentz contraction factor. To see how this comes 
about, we have to keep in mind that a thermometer which is kept at rest, by definition, 
can achieve equilibrium with respect to {\em energy} exchange with the system in
thermal contact with it, and not with momentum. In other words, the energies - that 
is the time components of the four-momentum - of the thermometer and the system will 
be shared (interaction energy is neglected) so as to maximize entropy, thus:
\beq\label{equi}
 \frac{\partial S}{\partial E}\Big|_{\rm T} = \frac{\partial S}{\partial E}\Big|_{\rm S}
\eeq
where S stands for system and T for thermometer. The left hand side, in the rest
frame of the thermometer, must be $1/T_{\rm T}$, i.e. the inverse temperature marked
by its gauge, while the right hand side is either $\gamma/T$ in the Einstein-Planck 
option (\ref{ep}) or $1/\gamma T$ in the alternative option (\ref{ovk}). 

Without delving the controversy in depth, my viewpoint is that the Einstein and Planck's 
- hence the most widely accepted in the past \cite{tolman} as well as today \cite{israel,hakim} 
- relativistic extension of the temperature concept is the correct one. If entropy 
is a Lorentz scalar, it must be a function of the invariant mass, that is 
$S=S(\sqrt{E^2 - {\bf P}^2})$. Hence:
$$
 \frac{\partial S}{\partial P^\mu} = \frac{\partial S}{\partial M} \frac{\partial}
{\partial P^\mu} \sqrt{E^2 - {\bf P}^2} = \frac{\partial S}{\partial M} \frac{P_\mu}{M}
 = \frac{\partial S}{\partial M} u_\mu
$$
The derivative of the entropy with respect to the mass of the system, that is its
rest energy, can be properly seen as the proper temperature, the one which would 
be measured by a thermometer at rest with the system, hence the above relation 
reads:
$$
 \frac{\partial S}{\partial P^\mu} = \frac{1}{T} u_\mu \equiv \beta_\mu
$$
where we have introduced the inverse temperature four-vector, or simply, the four-
temperature $\beta$, see Introduction. Hence, the entropy differential can be written as 
$$
 \di S = \frac{\partial S}{\partial P^\mu} \di P^\mu = \beta_\mu \di P^\mu 
 = \beta \cdot \di P
$$
which is apparently the Einstein-Planck extension (\ref{ep}). Instead, the alternative 
option suffers from a serious difficulty: to make sense of a differential relation
(\ref{ovk}), the four-momentum vector of a relativistic thermodynamic system must be 
a function of a scalar, the entropy. This is clearly counter-intuitive and against
any classical definition and experimental evidence, as entropy has to do only with 
the internal state of a system and should be independent of its collective motion. 
Therefore, the alternative by Ott and followers should be refused.

\section{Four-temperature $\beta$ and the $\beta$ frame}

The four-temperature $\beta$ is then the correct relativistic extension of the 
temperature notion. The four-temperature vector is ubiquitous in all relativistic 
thermodynamic formulae, such as the well known J\"uttner or Cooper-Frye distribution 
function:
$$
 f(x,p) = \frac{1}{\ee^{\beta \cdot p} \pm 1}
$$
Yet, $\beta$ is usually viewed as a secondary quantity obtained from previously defined
temperature and an otherwised defined velocity $u$, with $\beta=(1/T)u$. In this 
section, we will overturn this view.

One can make the definition of four-temperature operational, like in classical 
thermodynamics for the temperature, by defining an ideal "relativistic thermometer"
as an object able to instantaneously achieve equilibrium with respect to energy {\em 
and momentum} exchange. This implies that an ideal relativistic thermometer will 
istantaneously move at the same velocity as the system which is in contact with, 
besides marking its temperature, i.e. it will tell the $\beta$ vector in each spacetime 
point:
$$
 \frac{\partial S}{\partial P^\mu}_{\rm T} = \frac{\partial S}{\partial P^\mu}_{\rm S}
 \implies \beta^\mu_{\rm T} = \beta^\mu_{\rm S}
$$

Alternatively, one can retain the more traditional definition of thermometer, with 
an externally imposed four-velocity $u_{\rm T}$. In the latter case, going to the 
thermometer rest frame, one has, from the equation (\ref{equi}), the equality of the 
time components of the $\beta$ vectors in that frame:
$$
  \beta^0 = \beta^0_{\rm T}
$$
or
$$
  \beta \cdot u_{\rm T} = \frac{1}{T_{\rm T}}.
$$
Hence, a thermometer moving with four-velocity $u_{\rm T}$ in a system in local 
thermodynamical equilibrium, characterized by a four-vector field $\beta$, will 
mark a temperature:
\beq\label{temp}
 T_T = \frac{1}{\beta(x) \cdot v}.
\eeq
As the scalar product of two timelike unit vectors $ u \cdot v \ge 1$ and
$$
    u \cdot v = 1 \qquad {\rm iff} \;\; u = v
$$    
one has, according to (\ref{temp})
$$
  T_T \le T = \frac{1}{\sqrt{\beta^2}}  \qquad \qquad  T_T = T \qquad {\rm iff} 
  \;\;  u = u_{\rm T},
$$
that is the temperature marked by an idealized thermometer is maximal if it moves with
the same four-velocity of the (fluid) system. Thus:
$$
  T = 1/\sqrt{\beta^2}
$$
is the comoving, or proper, temperature.

Thereby, we can establish a thought operational procedure to {\em define} a four-velocity,
that is a frame, for a fluid based on the notion of local thermodynamical equilibrium 
at some spacetime point $x$:
\begin{itemize}  
\item{} put (infinitely many) ideal thermometers in contact with the relativistic 
system at the spacetime point $x$, each with a different four-velocity $u_{\rm T}$;
\item{} the ideal thermometer marking the {\em highest} temperature value $T$ moves, 
by construction, with the four-velocity $u(x) = T \beta(x) = 1/\sqrt{\beta^2} \beta(x)$.
\end{itemize}. 

We can thus define a four-velocity of a fluid just by using an ideal thermometer.
This makes the four-vector $\beta$ a more fundamental quantity than the fluid velocity. 
We defined this frame as $\beta$ frame \cite{betaframe} to distinguish it from the 
tradtional Landau and Eckart frames, from which it differs even in general global 
equilibrium states, as we explicitely showed in ref.~\cite{becaquantum}. The $\beta$
frame has many nice features and it is very convenient in general relativity, 
especially for quantum statistical mechanics, because at equilibrium it has a crucial 
feature: it is a {\em Killing vector field} as we will see in the next section.

\section{Quantum relativistic statistical mechanics at equilibrium}
\label{quantum}

In thermal quantum field theory, the usual task is to calculate mean values of physical 
quantities at thermodynamic equilibrium with an equilibrium density operator, whose
familiar form is:
\beq\label{homo1}
 \wrho = (1/Z) \exp [- \widehat H/T_0 + \mu_0 \wQ/T_0]
\eeq
where $T_0$ is the temperature and $\mu_0$ the chemical potential (the reason for
the 0 superscript will become clear soon) coupled to a conserved charge $\wQ$, and 
$Z$ the partition function. The above density operator can be obtained by maximizing 
the total entropy $S = -\tr (\wrho \log \wrho)$ with respect to $\wrho$ with the 
constraints of fixed total {\em mean} energy and fixed total {\em mean} charge. If 
a further constraint of fixed {\em mean} momentum vector is included, the density 
operator becomes manifestly covariant:
\beq\label{homo2}
 \wrho = (1/Z) \exp [- \beta \cdot \widehat P + \mu_0 \wQ/T_0]
\eeq
where $\widehat P$ is the four-momentum operator and $\beta$ is a four-vector
Lagrange multiplier for energy and momentum. The form (\ref{homo2}) is thus the
covariant form of (\ref{homo1}), which is a special case when $\beta=(1/T_0,{\bf 0})$.

However, the density operator (\ref{homo2}), is not the only form of global thermodynamic 
equilibrium, as one can add more constraints. For instance, one can include the
angular momentum and obtain \cite{landau,vilenkin}:
\beq\label{rotating}
  \wrho = (1/Z) \exp [-\widehat H /T_0 + \omega \widehat J_z/T_0 + \mu_0 \wQ/T_0]
\eeq
where $\widehat J_z$ is the angular momentum operator along some axis $z$, which 
represents a globally equilibrated spinning fluid with angular velocity $\omega$. 

The above (\ref{homo2}) and (\ref{rotating}) are indeed special cases of the most 
general thermodynamic equilibrium density operator, which can be obtained by maximizing 
the total entropy $S = -\tr (\wrho \log \wrho)$ with the constraints of given 
mean energy-momentum and charge {\em densities} at some specific time over some spacelike 
hypersurface $\Sigma$ \cite{weert,betaframe,nippon}. Therefore, the general equilibrium 
density operator can be written in a fully covariant form as \cite{zubarev,weert,weldon}: 
\beq\label{gencov}
  \wrho = (1/Z) \exp \left[- \int_\Sigma \di\Sigma_\mu  \left( \wT^{\mu\nu} \beta_\nu 
- \zeta \wj^\mu \right) \right]
\eeq
where $\wT^{\mu\nu}$ is the stress-energy tensor operator, $\wj^\mu$ a conserved current 
and $\zeta$ is a scalar whose meaning is the ratio between comoving chemical potential 
and comoving temperature. The four-vector field $\beta$ can be seen as a field of Lagrange
multipliers and no longer needs to be constant and uniform at equilibrium.

Indeed, for the right hand side of eq. (\ref{gencov}) to be a true, global equilibrium 
distribution, the integral must be independent of the particular $\Sigma$, which 
also means independent of time if $\Sigma$ is chosen to be $t=const$, as it was 
pointed out in Sect.~\ref{entrel}. Provided that the flux at some timelike boundary 
vanishes, this condition requires the divergence of the vector field in the integrand 
to be zero. If the stress-energy tensor $\wT$ and the current $\wj$ are covariantly
conserved, this requires $\zeta$ to be a constant scalar field and $\beta$ a Killing 
vector field, that is fulfilling the equation:
\beq\label{kill}
  \nabla_\mu \beta_\nu + \nabla_\nu \beta_\mu = 0  
\eeq
This condition for thermodynamic equilibrium has been known for a long time (see
e.g. \cite{degroot} for a kinetic derivation and \cite{becacov} for the above one).
The density operator (\ref{gencov}) is well suited to describe thermodynamic 
equilibrium in a general curved spacetime possessing a timelike Killing vector field.
It should be pointed out that extending the building blocks of quantum mechanics,
that is operators and Hilbert spaces, to curved spacetimes, features several major
difficulties, which can be partly circumvented by using the path integral formalism
\cite{birrel}. Thus, making full sense of expressions such as (\ref{gencov}) in 
curved spacetimes may not be trivial and it has been the subject of long discussion
and research which certainly goes beyond the scope of this work. Nevertheless, one 
can keep on using the operator formalism in an abstract algebraic sense, with the 
understood convention that traces are to be calculated by path integrals, so that 
the conclusion (\ref{kill}) holds. 

In Minkowski spacetime the general solution of the eq.~(\ref{kill}) is known:
\beq\label{killsol}
   \beta^\nu = b^\nu + \varpi^{\nu\mu} x_\mu
\eeq
where $b$ is a constant four-vector and $\varpi$ a constant antisymmetric tensor, 
which, because of eq.~(\ref{killsol}) can be written as an exterior derivative of the
$\beta$ field
\beq\label{thvort}
 \varpi_{\nu\mu} = -\frac{1}{2} (\partial_\nu \beta_\mu - \partial_\mu \beta_\nu)
\eeq
defined as {\em thermal vorticity}.
Hence, by using the eq.~(\ref{killsol}), the integral in eq.~(\ref{gencov}) can be 
rewritten as:
\beq\label{generatkill}
  \int_\Sigma \di\Sigma_\mu \; \wT^{\mu\nu} \beta_\nu = b_\mu {\wP}^\mu  
  - \frac{1}{2} \varpi_{\mu\nu} \wJ^{\mu\nu}
\eeq
and the density operator (\ref{gencov}) as:
\beq\label{gener2}
  \wrho = \frac{1}{Z} \exp \left[ - b_\mu {\wP}^\mu  
  + \frac{1}{2} \varpi_{\mu\nu} \wJ^{\mu\nu} + \zeta \wQ \right]
\eeq
where the $\wJ$'s are the generators of the Lorentz transformations:
$$
 \wJ^{\mu\nu} = \int_{\Sigma} \di \Sigma_\lambda \; \left( 
 x^\mu \wT^{\lambda\nu} - x^\nu \wT^{\lambda\mu} \right) 
$$
Therefore, besides the chemical potentials, the most general equilibrium density 
operator in Minkowski spacetime can be written as a linear combinations of the 10 
generators of its maximal continuous symmetry group, the orthocronous Poincar\'e 
group with 10 constant coefficients. 

It can be readily seen that the familiar density operator (\ref{homo1}) is obtained 
by setting $b=\frac{1}{T_0}(1,0,0,0)$ and $\varpi=0$, what we define as {\em homogeneous 
thermodynamic equilibrium}. The rotating global equilibrium in eq.~(\ref{rotating}) 
can be obtained as a special case of eq.~(\ref{gener2}) by setting:
\beq\label{rot}
 b_\mu = (1/T_0,0,0,0) \qquad \qquad \varpi_{\mu\nu} = (\omega/T_0) (g_{1\mu} g_{2\nu} 
- g_{1\nu} g_{2\mu})
\eeq
i.e. by imposing that the antisymmetric tensor $\varpi$ has just a ``magnetic"
part; thereby, $\omega$ gets the physical meaning of a costant angular velocity 
\cite{landau}. In fact, there is a third, not generally known, form which is conceptually
independent of the above two, which can be obtained by imposing that $\varpi$ has
just an "electric" (or longitudinal) part, i.e.:
\beq\label{acc}
 b_\mu = (1/T_0,0,0,0) \qquad \qquad \varpi_{\mu\nu} = (a/T_0) (g_{0\mu} g_{3\nu}
 - g_{3\mu} g_{0\nu})
\eeq
The resulting density operator is:
\beq\label{accdo}
  \wrho = (1/Z) \exp [-\widehat H /T_0 + a \widehat K_z/T_0]
\eeq
$\widehat K_z$ being the generator of a Lorentz boost along the $z$ axis. This 
represents a relativistic fluid with constant comoving acceleration along the $z$ 
direction. Note that the operators $\widehat H$ and $\widehat K_z$ are both 
conserved and yet, unlike in the rotation case (\ref{rotating}) they {\em do not}
commute with each other. This makes the density operator (\ref{accdo}) a very
peculiar kind of thermodynamic equilibrium \cite{becaprep}. 

\section{Killing vectors and Lie derivatives}

In this section I will prove a general property of any physical observable in general
thermodynamic equilibrium:\\

{\em The Lie derivative of any physical observable $X$ along the four-temperature vector
$\beta$ vanishes at thermodynamic equilibrium}\\

This statement makes it clear what thermodynamic equilibrium physically implies for 
an observer moving along a Killing vector field in a general spacetime, at it
will be discussed in the Sect.~\ref{conclu}.

A physical observable $X$ in quantum statistical mechanics is always defined as 
the mean value of a corresponding quantum operator, which can be either local or
resulting from an integration:
$$
  X = \tr (\wrho \widehat X(x)) 
$$
With a density operator given by (\ref{gencov}), the mean value will depend on the 
four-temperature field, the metric and $\zeta$, in a functional sense
$$
  X = X [\beta,\zeta,g]
$$
because so does the density operator $\wrho$. This is the most general dependence
that $X$ can have upon the data, i.e. the background metric and the thermodynamic fields
$\beta$ and $\zeta$ (in fact, the only non-trivial dependence will be on $\beta$ and
$g$ as $\zeta$ is constant at thermodynamic equilibrium). Expanding the functional
dependence, a local mean value $X$ will then depend, in general, on the derivatives of
all orders of both $\beta$ and $g$ calculated in $x$. Indeed, all the derivatives
at some point are what we need to know the supposedly analytic functions $\beta$ 
and $g$ in any other spacetime point. Furthermore, because of general covariance, 
we can choose an inertial set of coordinates in $x$ so that the first derivatives 
of the metric vanish, and all derivatives in $x$ of $\beta$ and $g$ at all orders 
in $x$ can be expressed as combinations of covariant derivatives of any order of 
$\beta$ and the Riemann tensor. In symbols:
\beq\label{xgen}
  X(x) = X [\beta,\zeta,g] = X (\beta,\nabla\beta,\nabla\nabla\beta,\ldots,g,
  R,\nabla R,\nabla\nabla R, \ldots)
\eeq
Indeed, being $\beta$ a Killing vector, it is known that its second covariant
derivative can be expressed as:
\beq\label{killrie}
  \nabla_\mu \nabla_\nu \beta_\lambda = R^\rho_{\; \mu\nu\lambda} \beta_\rho
\eeq
so that, effectively, the dependence on the four-temperature field at equilibrium
is just on the field and its covariant derivative. Therefore, the eq.~(\ref{xgen})
can be rewritten as:
\beq\label{xgen2}
  X(x) = X [\beta,\zeta,g] = X (\beta,\nabla\beta,g,R,\nabla R,\nabla\nabla R,\ldots)
\eeq

Altogether, $X$ can be seen as an analytic function of infinitely many arguments
and expanded in them. In general, the tensorial rank of $X$ determines how the 
arguments can appear in its expansion: for instance, if $X$ is a scalar, it will 
be expressed as all possible scalar combinations of the arguments with scalar coefficients 
depending on $\beta^2$, e.g.:
$$
 c_1(\beta^2) R^{\mu\nu\lambda\rho} R_{\mu\nu\lambda\rho} + c_2(\beta^2) R 
 + c_3(\beta^2) R^{\mu\nu} R_{\mu\nu} + c_4(\beta^2) 
 \nabla_\mu \beta_\nu \nabla^\mu \beta^\nu + \ldots
$$
where we have used the Ricci tensor and the curvature scalar.

A simple example of a relation (\ref{xgen2}) is the well known mean value of the 
stress energy at the homogeneous equilibrium in Minkowski spacetime with constant 
$\beta = b$ with $\varpi=0$ (see Section \ref{quantum}):
$$
  T^{\mu\nu}(x) = \frac{h(\beta^2)}{\beta^2} \beta^\mu\beta^\nu + p(\beta^2) g^{\mu\nu}
$$
where $h$ is the enthalpy density and $p$ the pressure, which are both functions
of $\beta^2$, i.e. the proper temperature. In curved spacetimes or in general 
equilibria in flat spacetime defined by the eq.~(\ref{killsol}), there can be much
more than the ideal form. Indeed, an expansion of the general relation (\ref{xgen2}) 
for the stress-energy tensor was envisaged in refs.~\cite{baier,roma} which was 
further studied and developed in several papers, e.g. refs.~\cite{kaminski,tifr,batta} 
with path integral methods; the coefficients of the expansion have been calculated 
in some relevant cases \cite{moore,becaquantum}. 

Therefore, in order to prove the statement at the beginning of this section, we 
just need to show that any argument of $X$ in the eq.~(\ref{xgen}) has a vanishing
Lie derivative along $\beta$. For $\beta$ this is trivial, for $g$ it is true by
definition of Killing vector, i.e. the eq.~(\ref{kill}) itself. To proceed and 
show that this holds for any other argument, we need first to prove the following:

\begin{prop}
{\it For any vector field $V$, the Lie derivative along a Killing field $\beta$
commutes with the covariant derivative, that is ${\cal L}_\beta (\nabla V) = 
\nabla {\cal L}_\beta(V)$}
\end{prop}

To show this, we expand the Lie derivative definition:
\beq\label{lie1}
  \lieb(\nabla_\mu V_\nu) = \beta^\lambda \nabla_\lambda \nabla_\mu V_\nu
 + \nabla_\mu \beta^\lambda \nabla_\lambda V_\nu + \nabla_\nu \beta^\lambda 
  \nabla_\mu V_\lambda
\eeq
Now we use the commutator of two covariant derivatives:
\beq\label{commrel}
  \nabla_\lambda \nabla_\mu V_\nu - \nabla_\mu \nabla_\lambda V_\nu = 
  R^\rho_{\; \nu\mu\lambda} V_\rho
\eeq
for the first term on the right hand side of (\ref{lie1}), and the Leibniz rule
for the covariant derivative of the other two terms. Hence:
\bea\label{lie2}
 \lieb(\nabla_\mu V_\nu) = && \beta^\lambda \nabla_\mu \nabla_\lambda V_\nu
  + \beta^\lambda R^\rho_{\; \nu\mu\lambda} V_\rho + \nabla_\mu (
 \beta^\lambda \nabla_\lambda V_\nu) \nonumber \\
 && - \beta^\lambda \nabla_\mu \nabla_\lambda V_\nu
 + \nabla_\mu (\nabla_\nu \beta^\lambda V_\lambda) -\nabla_\mu \nabla_\nu 
  \beta^\lambda V_\lambda \nonumber \\
 = && \beta^\lambda R^\rho_{\; \nu\mu\lambda} V_\rho -\nabla_\mu \nabla_\nu 
  \beta^\lambda V_\lambda + \nabla_\mu \lieb(V_\nu)
\eea
where we have again used the Lie derivative definition, for a vector field. Now
we can use the eq.~(\ref{killrie}), so that the first two terms on the right hand
side of eq.~(\ref{lie2}) cancel:
\begin{eqnarray*}
 && \beta^\lambda R^\rho_{\; \nu\mu\lambda} V_\rho -\nabla_\mu \nabla_\nu 
  \beta^\lambda V_\lambda = R_{\rho\nu\mu\lambda} V^\rho \beta^\lambda -
  R_{\rho\mu\nu\lambda} \beta^\rho V^\lambda \nonumber \\ &&
 = (R_{\lambda\nu\mu\rho} - R_{\rho\mu\nu\lambda})
  V^\lambda \beta^\rho = (R_{\mu\rho\lambda\nu} - R_{\rho\mu\nu\lambda}) 
 V^\lambda \beta^\rho \nonumber \\
 &&  = (R_{\rho\mu\nu\lambda} - R_{\rho\mu\nu\lambda}) V^\lambda \beta^\rho = 0 
\end{eqnarray*}
where we have used the symmetry properties of the Riemann tensor indices. Thus,
eq.~(\ref{lie2}) yields the sought relation:
$$
  \lieb(\nabla_\mu V_\nu) = \nabla_\mu \lieb(V_\nu)
$$
and this concludes the proof.
\\

By using the Leibniz rule for the covariant derivative of a tensor product, it is 
straightforward to extend the above relation to the Lie derivative of any tensor field 
$T$, that is:
\beq\label{lie3}
  \lieb(\nabla T) = \nabla \lieb(T)
\eeq
A straightforward consequence of the above relation is that $\lieb (\nabla \beta)=0$
being $\lieb (\beta) = 0$. 

The last step to prove the initial statement is to show that the Riemann tensor 
has vanishing Lie derivative along $\beta$, that is:
\beq\label{lierz}
 \lieb (R) = 0
\eeq
which implies, in view of the (\ref{lie3}) that all Lie derivatives of $\nabla R,
\nabla\nabla R,\ldots$ vanish. Let us take an arbitrary vector field $V$ and write
the Lie derivative of the (\ref{commrel}):
$$
 \lieb (\nabla_\lambda \nabla_\mu V_\nu - \nabla_\mu \nabla_\lambda V_\nu) = 
  \lieb (R^\rho_{\; \nu\mu\lambda} V_\rho) 
$$
By using Leibniz rule and the (\ref{lie3}) we get:
$$
  (\nabla_\lambda \nabla_\mu -\nabla_\mu \nabla_\lambda) \lieb(V)_\nu
  = \lieb (R^\rho_{\; \nu\mu\lambda}) V_\rho +  R^\rho_{\; \nu\mu\lambda}
  \lieb(V_\rho)
$$
By using again the (\ref{commrel}) for the left hand side:
$$
  R^\rho_{\; \nu\mu\lambda} \lieb(V_\rho) = \lieb (R^\rho_{\; \nu\mu\lambda}) V_\rho 
+  R^\rho_{\; \nu\mu\lambda} \lieb(V_\rho)
$$
whence we conclude that:
$$
\lieb (R^\rho_{\; \nu\mu\lambda}) V_\rho = 0 
$$
for any vector field $V$. Thus, we obtain the (\ref{lierz}), which finally demonstrate
the general statement at the beginning of the section.

\section{Concluding remarks}
\label{conclu}

The general stationarity equation implied by the vanishing of the Lie derivative
for a scalar reads:
$$
 \lieb (S) = \beta^\lambda \partial_\lambda S = \sqrt{\beta^2} 
 u^\lambda \nabla_\lambda S \equiv \sqrt{\beta^2} D S = 0
$$
implying that a scalar quantity does not change along the $\beta$ flow. That is, 
a comoving observer with four-velocity $u = \beta/\sqrt{\beta^2}$ will measure the 
same temperature, energy density, pressure and any other scalar field.

Instead, for a vector field:
\beq\label{vect}
 \lieb (V_\mu) = \beta^\lambda \nabla_\lambda V_\mu + (\nabla_\mu \beta_\lambda) 
 V^\lambda = 0
\eeq
As $\beta$ is a Killing vector, its covariant derivative is antisymmetric, thus,
one can extend the (\ref{thvort}) to the general relativistic case, that is
$\varpi_{\mu\lambda} = - \nabla_\mu \beta_\lambda$. If one sets:
$$
  \Omega_{\mu\lambda} \equiv \frac{1}{\sqrt{\beta^2}} \varpi_{\mu\lambda}  
$$
the $\Omega$ is an antisymmetric tensor such that, according to (\ref{vect}):
\beq\label{vect2}
 D V_\mu = \Omega_{\mu\lambda} V^\lambda
\eeq
These are the well known (in general relativity) equations of motion of an orthonormal 
tetrad frame, the relativistic extension of the classical Poisson equations for the 
motion of a rigid frame. The consequence of (\ref{vect2}), for a vector field $V$ at 
thermodynamic equilibrium, is that its components are constant for a comoving observer 
only if he has an associated tetrad frame - which must include the normalized Killing 
vector itself as time direction - which is Lie-transported, that is with vanishing 
Lie derivative; the same holds for any tensor field. 

It is also worth pointing out another interesting consequence of this formulation
of general relativistic thermodynamics:\\

{\it A free-falling ideal thermometer in a fluid at global thermodynamic equilibrium
will mark a constant temperature $T_{\rm T} = 1/(\beta \cdot u)$ with $\beta$ the
four-temperature of the fluid and $u$ the four-velocity of the thermometer}
\\

This is a straightforward consequence of the well known conservation theorem 
for a geodesic motion in spacetimes with Killing vectors \cite{misner} and the 
equation (\ref{temp}).

\section*{Acknowledgments}

I am grateful to D. Seminara for interesting and clarifying discussions.





\begin{thebibliography}{99}

\bibitem{einstein}
 A.~Einstein, Jahrb. Radioakt. Elektron., {\bf 4} (1907) 411.

\bibitem{planck}
 M.~Planck, Ann. Phys. (Liepzig) {\bf 26} (1908) 1.

\bibitem{ott} 
 H.~Ott, Zeit. Phys. {\bf 175} (1963) 70.

\bibitem{kampen}
  N.~G.~van Kampen, Phys.\ Rev.\  {\bf 173} (1968) 295.

\bibitem{arabi}
 M.~Khaleghy, F.~Qassemi, arXiv preprint physics/0506214 (2005).


\bibitem{tolman}
 R.~C.~Tolman, {\it Relativity, Thermodynamics and Cosmology}, Clarendon Oxford
  (1934).

\bibitem{israel} 
 W.~Israel, Annals Phys.\  {\bf 100}, 310 (1976); W.~Israel and J.~M.~Stewart,
  Annals Phys.\  {\bf 118}, 341 (1979).

\bibitem{hakim}
 R.~Hakim, {\it Introduction to Relativistic Statistical Mechanics}, World Scientific
 (2011).
  
\bibitem{betaframe}
  F.~Becattini, L.~Bucciantini, E.~Grossi and L.~Tinti,  Eur.\ Phys.\ J.\ C {\bf 75} 
  (2015) 191.

\bibitem{becaquantum}
  F.~Becattini and E.~Grossi, Phys.\ Rev.\ D {\bf 92}, no. 4, 045037 (2015).

\bibitem{landau}
  L.~Landau, L.~Lifshitz, {\it Statistical Physics}, Pergamon Press (1980).
  
\bibitem{vilenkin}
 A. Vilenkin, Phys. Rev. D {\bf 21} 2260 (1980).

\bibitem{weert}
 Ch.~G.~Van~Weert, Ann.\ Phys.\ {\bf 140}, 133 (1982).  
 
\bibitem{nippon}
  T.~Hayata, Y.~Hidaka, T.~Noumi and M.~Hongo, Phys.\ Rev.\ D {\bf 92}, no. 6, 
  065008 (2015).  
 
\bibitem{zubarev} 
 D.~N.~Zubarev, A.~V.~Prozorkevich, S.~A.~Smolyanskii, Theoret. and Math. 
 Phys. 40 (1979), 821. 

\bibitem{weldon}
  H.~A.~Weldon, Phys.\ Rev.\ D {\bf 26}, 1394 (1982).
 
\bibitem{degroot}
  S.~R.~De Groot, W.~A. van Leeuwen, Ch.~G. van Weert, {\it Relativistic
  kinetic theory}, North Holland (1980). 

\bibitem{becacov} 
  F.~Becattini, Phys.\ Rev.\ Lett.\  {\bf 108} (2012) 244502.

\bibitem{birrel}
  N.~C.~Birrell, P.~C.~W.~Davies, {\it Quantum fields in curved space}, Cambridge 
  Monographs on Mathematical Physics (1984).

\bibitem{becaprep}
  F. Becattini, in preparation.
  
\bibitem{baier}
  R.~Baier, P.~Romatschke, D.~T.~Son, A.~O.~Starinets and M.~A.~Stephanov, 
  JHEP {\bf 0804}, 100 (2008).

\bibitem{roma}
  P.~Romatschke, Class.\ Quant.\ Grav.\  {\bf 27}, 025006 (2010).
 
\bibitem{kaminski}
  K.~Jensen, M.~Kaminski, P.~Kovtun, R.~Meyer, A.~Ritz and A.~Yarom,
  Phys.\ Rev.\ Lett.\  {\bf 109}, 101601 (2012).
  
\bibitem{tifr} 
  N.~Banerjee, J.~Bhattacharya, S.~Bhattacharyya, S.~Jain, S.~Minwalla and T.~Sharma,
  JHEP {\bf 1209}, 046 (2012).
 
\bibitem{batta}
  S.~Bhattacharyya, JHEP {\bf 1207}, 104 (2012).

\bibitem{moore}  
  G.~D.~Moore and K.~A.~Sohrabi, JHEP {\bf 1211}, 148 (2012).

\bibitem{misner}
 C.~W.~Misner, K.~S.~Thorne, J.~A.~Wheeler, {\it Gravitation} (1973)
 W.~H.~Freeman and Company.
   
\end{thebibliography}
\end{document}